\documentclass[amssymb,aps, prl,10pt]{revtex4-1}

\usepackage{amsmath}
\usepackage{fullpage}
\usepackage{amssymb}
\usepackage{amsfonts}
\usepackage{color}
\usepackage[pdftex]{graphicx}
\usepackage{epstopdf}
\usepackage{soul}
\usepackage{xcolor}
\usepackage{multirow}

\definecolor{LightGreen}{RGB}{144, 238, 144}
\sethlcolor{LightGreen}

\providecommand{\e}[1]{\ensuremath{\times 10^{#1}}}
\newcommand{\Emu}{\mathbf{E_{\mu}}} 
\newcommand{\Hmu}{\mathbf{H_{\mu}}} 
\newcommand{\Eint}{\mathbf{E^{\mathrm{int}}}} 
\newcommand{\Es}{\mathbf{E^{\mathrm{s}}}} 
\newcommand{\Ed}{\mathbf{E^{\mathrm{d}}}} 
\newcommand{\Hs}{\mathbf{H^{\mathrm{s}}}} 
\newcommand{\Hint}{\mathbf{H^{\mathrm{int}}}} 
\newcommand{\Ae}{\mathbf{A}_{j,m_z}^{(e)}}
\newcommand{\Ex}{\mathbf{E}_{j,m_z,q}^{(x)}}

 \newcommand{\Am}{\mathbf{A}_{j,m_z}^{(m)}} 
\newcommand{\dd}{\mathrm{d}}
\newcommand{\Vmu}{V_{\mu}}
\newcommand{\Qmu}{Q_{\mu}}

\newcommand{\omu}{\omega_{\mu}}
 
\newcommand{\xhat}{\hat{\mathbf{x}}}
\newcommand{\rhat}{\hat{\mathbf{r}}}
\newcommand{\that}{\hat{\mathbf{\theta}}}
\newcommand{\phat}{\hat{\mathbf{\phi}}}
\newcommand{\yhat}{\hat{\mathbf{y}}}
\newcommand{\zhat}{\hat{\mathbf{z}}}
\newcommand{\G}{\overleftrightarrow{\mathbf{G}}^{(m)}} 
\newcommand{\Ge}{\overleftrightarrow{\mathbf{G}}^{(e)}} 
\newcommand{\Yjm}{Y_{j,m_z}}

\newcommand{\Rje}{R_j(k_{j,q}^{(e)})}
\newcommand{\Rjx}{R_j(k_{j,q}^{(x)})}
\newcommand{\Ve}{V_{j,m_z,q}^{(e)}}
\newcommand{\Vx}{V_{j,m_z,q}^{(x)}}
\newcommand{\Vm}{V_{j,m_z,q}^{(m)}}
\newcommand{\ke}{k_{j,q}^{(e)}}
\newcommand{\xe}{x_{j,m_z,q}^{(e)}}
\newcommand{\xx}{x_{j,m_z,q}^{(x)}}
\newcommand{\xm}{x_{j,m_z,q}^{(m)}}
\newcommand{\km}{k_{j,q}^{(m)}}
\newcommand{\kx}{k_{j,q}^{(x)}}
\newcommand{\ox}{\omega_{j,m_z,q}^{(x)}}
\newcommand{\ome}{\omega_{j,q}^{(e)}}
\newcommand{\omm}{\omega_{j,q}^{(m)}}
\newcommand{\omx}{\omega_{j,q}^{(x)}}
\newcommand{\Pjm}{P_{j}^{\vert m_z\vert}(\cos\theta)}
\newcommand{\ace}{a^d_E(j,m_z)}
\newcommand{\acm}{a^d_M(j,m_z)}
\newcommand{\aVx}{\left\langle \Vx \right\rangle}
\newcommand{\aVm}{\left\langle \Vm \right\rangle}
\newcommand{\aVe}{\left\langle \Ve \right\rangle}

\begin{document}

\title{Purcell factor of Mie resonators featuring electric and magnetic modes}

\author{Xavier Zambrana-Puyalto}
\email{xavier.zambrana@fresnel.fr}
\affiliation{Aix-Marseille Universit\'{e}, CNRS, Centrale Marseille, Institut Fresnel UMR 7249, 13013 Marseille, France}
\author{Nicolas Bonod}
\email{nicolas.bonod@fresnel.fr}
\affiliation{Aix-Marseille Universit\'{e}, CNRS, Centrale Marseille, Institut Fresnel UMR 7249, 13013 Marseille, France}

\begin{abstract}
We present a modal approach to compute the Purcell factor in Mie resonators exhibiting both electric and magnetic resonances. The analytic expressions of the normal modes are used to calculate the effective volumes. We show that important features of the effective volume can be predicted thanks to the translation-addition coefficients of a displaced dipole. Using our formalism, it is easy to see that, in general, the Purcell factor of Mie resonators is not dominated by a single mode, but rather by a large superposition. Finally we consider a silicon resonator homogeneously doped with electric dipolar emitters, and we show that the average electric Purcell factor dominates over the magnetic one. 
\end{abstract}

\maketitle

\section{Introduction}
Dielectric Mie resonators made of moderate refractive index materials, such as semi-conductor materials in the optical or near infra-red regime, can feature electric and magnetic modes of low order. These modes can be easily and efficiently excited either from near or far field illumination \cite{Evlyukhin2010,Evlyukhin11,GarciaEtxarri11}. This novel class of photonic resonators is very promising to design directive antennas \cite{Rolly12c,Rolly2013,Krasnok14}, to enhance the electric or magnetic near field intensities \cite{Sigalas2007,Albella2013,Boudarham2014,Zhang2014}, to design subwavelength sized light cavities \cite{Rolly2012,Schmidt12,Albella2013}, to host frequency mixing processes such as third harmonic generation \cite{Shcherbakov2014}, and even to create isotropically polarized speckle patterns \cite{preSchmidt2014}. 
Contrarily to spherical nano-metallic particles, dielectric Mie resonators exhibit both electric and magnetic resonant modes \cite{Zhao09,Evlyukhin11,GarciaEtxarri11}. However, it is still unclear whether magnetic modes are a good platform to enhance light-matter interaction in silicon subwavelength-sized cavities. An interesting way to address this question is to study the Purcell factor of these cavities.

The Purcell factor is a figure of merit widely used to characterize light-matter interaction in photonic cavities \cite{Purcell1946}. In quantum electrodynamics, for example, it has been widely used to study Rabi oscillations in the weak-coupling regime \cite{Gerard2003}. The Purcell factor quantifies the enhancement of the spontaneous decay rate of a dipolar emitter coupled with a cavity, and it is defined as \cite{Purcell1946,Gerard2003}: 
\begin{equation}
\label{PF}
F_P(\omega)=\dfrac{6\pi c^3 Q_{\mu}}{\omega^3 \Vmu} \sim \dfrac{\Gamma}{\Gamma_0}
\end{equation}
with $\omega$ being the frequency of the dipole; $c$ the speed of light; $\Qmu$ the quality factor of the electromagnetic mode of the cavity ($\Emu$) that the emitter is mainly coupled to; $\Vmu$ the effective volume of the mode $\Emu$; and $\Gamma$ ($\Gamma_0$) the spontaneous decay rate of a dipolar emitter in the presence of an optical cavity (vacuum).

The design of photonic cavities at the nanoscale brought the attention of the nano-optics community to the Purcell factor \cite{Agio2013}. Nevertheless, the first attempts of using the Purcell factor given by eq.(\ref{PF}) to compute the enhancement of spontaneous decay rates ($\Gamma/\Gamma_0$) were not as accurate as expected. It was seen that the Dyadic Green tensor formulation \cite{Novotny2006} used to compute $\Gamma/\Gamma_0$ did not yield the same results as the Purcell factor given by eq.(\ref{PF}). As noted by Koenderink \cite{Koenderink2010}, the use of eq.(\ref{PF}) to compute spontaneous decay rates is based on a few restrictive assumptions, which are not usually fulfilled by photonic nano-cavities. Hence, some changes needed to be done in the Purcell factor formula so that $F_P(\omega)=\Gamma / \Gamma_0$. These changes can be summarized in two: extension of the Purcell factor to a scenario where the dipolar emitter couples to a superposition of modes $\Emu$ of the cavity; and redefinition of the effective volume $\Vmu$ of each of the $\Emu$ normal modes of the resonator. One of the first attempts to redefine the effective volume for spherical cavities was proposed by Colas des Francs \textit{et al.} in 2012 \cite{Derom2012,Colas2012}. The proposed method used the equality between the classical normalized decay rates \cite{Chew1976,Ruppin1982,Chew1987} and the Purcell factor given by eq.(\ref{PF}). That is, the effective volume $\Vmu$ was computed by substitution:
\begin{equation}
F_P(\omega)=\dfrac{6\pi c^3 Q_{\mu}}{\omega^3 \Vmu} \equiv \dfrac{\Gamma}{\Gamma_0} \Longrightarrow \Vmu = \dfrac{6\pi c^3 \Qmu }{\omega^3} \dfrac{\Gamma_0}{\Gamma}
\end{equation}
With this empirical formulation, an effective volume is obtained for each multipolar order of the spherical resonator. Recently, attention was focused on the definition of the effective volume with respect to the normal modes of arbitrary resonators \cite{Sauvan2013,Kristensen2013,Muljarov2014}. The normal modes of the system are solutions of Maxwell equations in the absence of sources, with outgoing radiation conditions, and with a certain normalization \cite{Kristensen2012,Sauvan2013,Ge2014,Muljarov2014}. Due to the outgoing radiation conditions, the Hamiltonian of the system is non-hermitian and the normal modes of the system have complex eigenfrequencies $\omu$ \cite{Sauvan2013,Grigoriev2013,Vial2014}. Let us notice that a normalization condition of the kind $\int_V \vert \Emu(\mathbf{r},\omu) \vert^2 \dd V$ cannot be applied to normal modes with resonant complex frequencies $\omu$ because $\Emu(\vert \mathbf{r} \vert \rightarrow \infty, \omu)\rightarrow \infty$ \cite{Koenderink2010,Kristensen2012,Kristensen2013,Sauvan2013,Bai2013,Muljarov2014}. In this work, we define the normal modes associated to the eigenfrequencies $\omu$ following the normalization condition given in \cite{Muljarov2014,Doost2014}:
\begin{equation}
1=\int_V \mathbf{E_{\mu}} \cdot \epsilon(\mathbf{r}) \Emu \dd \mathbf{r} + \dfrac{c^2}{2\omu^2} \oint_{\partial V} \left( \Emu \cdot \dfrac{\partial}{\partial s}r \dfrac{\partial  \mathbf{E_{\mu}}}{\partial r}  - r \dfrac{\partial  \mathbf{E_{\mu}}}{\partial r} \cdot \dfrac{\partial \Emu}{\partial s} \right) \dd \mathcal{S}
\label{NM}
\end{equation}
where all the magnitudes that depend on the frequency are particularized at $\omega = \omu$. At this point, two comments are in order. Firstly, the normalization condition given by eq.(\ref{NM}) can also be expressed as \cite{Muljarov2014,Sauvan2013}:
\begin{equation}
1 = \int_V \left[ \Emu \cdot \epsilon(\mathbf{r}) \Emu - \Hmu \cdot   \mu_0 \Hmu  \right] \dd \mathbf{r}
\end{equation}
As Sauvan \textit{et al.} showed in \cite{Sauvan2013}, this type of normalization is especially well-suited for numerical calculations. The underlying reason is that the volume integrals can be numerically evaluated using a perfectly-matched-layer technique. Furthermore, their normalization condition is easily extended for dispersive and magnetic materials  \cite{Sauvan2013}:
\begin{equation}
1 = \int_V \left[ \Emu \cdot \dfrac{\partial (\omega \epsilon)}{\partial \omega} \Emu - \Hmu \cdot   \dfrac{\partial (\omega \mu)}{\partial \omega} \Hmu  \right] \dd \mathbf{r}
\end{equation}
Secondly, the basis of normal modes $(\Emu,\Hmu)$ for leaky cavities with arbitrary geometries is not complete, in general \cite{Kristensen2013}. However, for spherical leaky cavities, it has been demonstrated that the multipolar fields are a complete basis to describe the problem \cite{Leung1996,Lee1999}.

\section{General expressions for electric and magnetic emitters}
The recent advances on the modal analysis of Purcell factors were performed with electric dipolar sources. The formulation has been presented in \cite{Kristensen2013,Sauvan2013,Muljarov2014}, where the enhancement of the decay rate was computed as:
\begin{eqnarray}
F(\omega)& = &\dfrac{6\pi c}{n \omega} \mathrm{Im} \left\lbrace \mathbf{p}^* \cdot \Ge(\mathbf{r_0},\mathbf{r_0}) \cdot \mathbf{p} \right\rbrace = \dfrac{3\pi c^3}{\omega} \sum_{\mu} \mathrm{Im} \left\lbrace \dfrac{1}{\Vmu \omu (\omu - \omega)} \right\rbrace 
\label{Fe}
\end{eqnarray}
with 
\begin{equation}
\Vmu = \dfrac{1}{\left( \mathbf{u_p} \cdot \mathbf{E_{\mu}(\mathbf{r_p},\omu)}  \right)^2   }
\label{Ve}
\end{equation}
where $\mathbf{p}=p \mathbf{u_p}$ is the electric dipole moment of the emitter, and $\mathbf{u_p}$ is a unitary vector. However, interest in magnetic spontaneous emission has been growing over the last years, in particular using trivalent lanthanide ions \cite{Noginova2009,Karaveli11,Aigouy14}. This strong interest in coupling lanthanide ions to nano-cavities makes us emphasize the fact that both electric and magnetic emitters can give rise to an enhanced spontaneous decay rate. Following the formalism shown in \cite{Muljarov2014}, next we give the Purcell factor formula for an emitting magnetic dipole. The interaction Hamiltonian for a dipolar magnetic interaction is \cite{Grynberg2010}:
\begin{equation}
H_I=-\mathbf{m} \cdot \mathbf{B}
\end{equation}
where $\mathbf{m}=m \mathbf{u_m}$ is the magnetic dipole moment of the emitter, and $\mathbf{u_m}$ is a unitary vector. If $\G$ is the magnetic dyadic Green function \cite{Narayanaswamy2010}, then the Purcell factor can be computed as:
\begin{eqnarray}
\label{PFg}
F(\omega)& = &\dfrac{6\pi c}{n \omega} \mathrm{Im} \left\lbrace \mathbf{m}^* \cdot \G(\mathbf{r_0},\mathbf{r_0}) \cdot \mathbf{m} \right\rbrace = \dfrac{3\pi c^3}{\omega} \sum_{\mu} \mathrm{Im} \left\lbrace \dfrac{1}{\Vmu \omu (\omu - \omega)} \right\rbrace 
\end{eqnarray}
where 
\begin{equation}
\label{Vmag}
\Vmu = \dfrac{1}{\left( \mathbf{u_m} \cdot \mathbf{B_{\mu}(\mathbf{r_p},\omu)}  \right)^2   }
\end{equation}
It is important to note that eqs.(\ref{Fe}) and (\ref{PFg}) cannot be directly added. That is, the enhancement of decay rates of a dipolar emitter which is a linear combination of an electric and a magnetic dipole ($\mathbf{e}=\alpha \mathbf{p}+ \beta \mathbf{m}$) is computed with the following formula \cite{Novotny2006}:
\begin{equation}
\dfrac{P}{P_0} = -\dfrac{1}{2P_0} \int_V \mathrm{Re} \left\lbrace \mathbf{j}^* \cdot \mathbf{E}  \right\rbrace \dd V 
\label{Power}
\end{equation}
where $\mathbf{j}= - i \omega \alpha \mathbf{p} \delta(\mathbf{r} - \mathbf{r_p}) + \nabla \times \beta \mathbf{m} \ \delta(\mathbf{r} - \mathbf{r_p})$ \cite{Hnizdo2012,Poddubny2013}, and $P_0$ is given in \cite{Novotny2006}. Using the currents produced by both dipoles, $P$ can be written as:
\begin{equation}
P=\dfrac{\omega}{2} \mathrm{Im} \left\lbrace \alpha^* \mathbf{p}^* \cdot \mathbf{E}(\mathbf{r_0}) + \beta^* \mathbf{m}^* \cdot \mathbf{B}(\mathbf{r_0})  \right\rbrace
\label{Pem}
\end{equation}
where $\mathbf{E}(\mathbf{r_0})$ and $\mathbf{B}(\mathbf{r_0})$ can be computed using the Green tensor formalism \cite{Novotny2006}. Notice that when the emitter is a superposition of an electric and a magnetic dipole, $\mathbf{E}(\mathbf{r_0}) \neq \Ge(\mathbf{r_0},\mathbf{r_0}) \cdot \mathbf{p} $ since:
\begin{equation}
\mathbf{E}(\mathbf{r_0})= i \omega \mu \mu_0 \int_V \Ge(\mathbf{r},\mathbf{r_0}) \cdot \mathbf{j}(\mathbf{r}) \dd V
\end{equation}
with $\mathbf{j}(\mathbf{r})$ given by the equation above.

\section{Multipolar expression of the normal modes}
Here, thanks to the analytic formulation of Mie theory, we analytically compute the normal modes of the system and their associated normal frequencies. In particular, for a Mie resonator, due to the symmetries of the system \cite{ZambranaThesis}, the normal modes of the system are the multipolar fields, $\Emu=\mathbf{E}_{j,m_z,q}^{(x)}$, where $j=1,..,\infty$, $m_z=-j,...,j$, $q=1,..,\infty$ and $(x)=(e),(m)$. 
The meaning of $j,m_z,$ and $(x)$ is directly related to the following differential operators \cite{Rose1955,Tung1985,ZambranaThesis}:
\begin{eqnarray}
J^2 \left[ \Ex  \right] &=& j(j+1) \Ex \\
J_z  \left[ \Ex  \right] &=& m_z \Ex \\
\Pi  \left[ \Ex  \right] &=& \left\lbrace \begin{array}{l}
(-)^j\mathbf{E}_{j,m_z,q}^{(m)} \\
(-)^{j+1}\mathbf{E}_{j,m_z,q}^{(e)}
\end{array}
\right.
\end{eqnarray} 
Nonetheless, $q$ is different in this sense. Given a fixed combination of $\left\lbrace j',m'_z,(x') \right\rbrace$, there exists a numerable infinite of $q$ values such that $\mathbf{E}_{j',m'_z,q}^{(x')}$ is a normal mode of Maxwell equations with an associated eigenfrequency $\omu=\omega_{j',m'_z,q}^{(x')}$ \cite{Oraevsky2002,Derom2012}. These eigenfrequencies $\omega_{j',m'_z,q}^{(x')}$ can be found by solving the following two transcendental equations:
\begin{align}
\label{oe}
n_r^2 j_j(n_r \xe )[\xe  h_j^{(1)}(\xe )]' & \ \ = & h_j^{(1)}(\xe ) [n_r \xe  j_j(n_r \xe )]' & \quad \text{for} & (x)=(e)  \\
j_j(n_r \xm)[\xm h_j^{(1)}(\xm)]' & \ \ = & h_j^{(1)}(\xm) [n_r \xm j_j(n_r \xm)]'& \quad \text{for} & (x)=(m) 
\label{om}
\end{align}
where $\xx = \frac{ R}{c} \ox $.

Here, it is important to note a few facts. Firstly, eqs.(\ref{oe}-\ref{om}) do not depend on $m_z$, therefore $\ox=\omega_{j,q}^{(x)}$. Secondly, in eqs.(\ref{oe}-\ref{om}), the dependence on $j$ is explicit on the spherical Bessel functions of the different kinds, and the dependence on $(x)$ is apparent as the equations for electric and magnetic frequencies are different. However, the dependence on $q$ is not explicit. That is because, given a fixed $j'$, $q$ is an integer that numerates all the solutions to eqs.(\ref{oe}-\ref{om}) in ascending order, \textit{i.e.} $\mathrm{Re}\left\lbrace \omega_{j',1}^{(x)} \right\rbrace < \mathrm{Re}\left\lbrace \omega_{j',2}^{(x)} \right\rbrace < \mathrm{Re}\left\lbrace \omega_{j',3}^{(x)} \right\rbrace< ...$. 

Due to the normalization condition imposed by eq.(\ref{NM}), the normal modes of the system are \cite{Doost2014,Muljarov2014}:
\begin{eqnarray}
\label{Ejme}
\mathbf{E}_{j,m_z,q}^{(m)} (r,\theta,\phi)& = & A_j^{(m)} R_j(k_{j,q}^{(m)})\left[ \dfrac{1}{\sin \theta} \dfrac{\partial Y_{j,m_z}}{\partial \phi}  \that -\dfrac{\partial \Yjm}{\partial \theta} \phat \right] \\
\mathbf{E}_{j,m_z,q}^{(e)} (r,\theta,\phi) & =& \dfrac{A_j^{(e)}}{\epsilon(r)k_{j,q}^{(e)}r} \left[ j(j+1) R_j(k_{j,q}^{(e)}) \Yjm \rhat + \right. \\
 & & \left. \dfrac{\partial \left( r \Rje  \right)}{\partial r}  \dfrac{\partial \Yjm}{\partial \theta}\that + \dfrac{\partial \left( r \Rje   \right)}{\partial r} \dfrac{1}{\sin\theta} \dfrac{\partial\Yjm}{\partial \phi} \phat \right] \nonumber
 \label{Ejmm}
\end{eqnarray}
where $(r,\theta,\phi)$ are the spherical coordinates in the real space; $\km$ and $\ke$ are related to the eigenfrequencies of the mode, $\kx = \omega_{j,q}^{(x)}/ c$ with $c$ being the speed of light in vacuum;
\begin{eqnarray}
A_j^{(m)}& = &\sqrt{\dfrac{2}{j(j+1)R^3(n_r^2-1)}} \\
A_j^{(e)}& =&n_r^2 A^{(m)} \left( \sqrt{ \left[ \dfrac{j_{j-1}(n_r\ke R)}{j_j(n_r\ke R)} -  \dfrac{j}{n_r \ke R } \right] + \dfrac{j(j+1)}{(k_{j,q}^{(e)}R)^2}}   \right)^{-1}
\end{eqnarray}
are normalization constants, with $j_j(x)$ being the spherical Bessel function; 
\renewcommand{\arraystretch}{1.5}
\begin{equation}
\Rjx = \left\lbrace  \begin{array}{lcc}
\dfrac{j_j(n_r \kx r)}{j_j(n_r \kx R)} & \text{for} & r \leq R \\
\dfrac{h_j(\kx r)}{h_j(\kx R)} & \text{for} & r > R
\end{array} \right.
\end{equation}
\renewcommand{\arraystretch}{1}
where $h_j(x)$ is a spherical Hankel function of the first kind;
\begin{equation}
\Yjm = \sqrt{ \dfrac{2j+1}{2} \dfrac{(j-\vert m_z \vert)!}{(j+\vert m_z \vert)!}  } P_{j}^{\vert m_z\vert}(\cos\theta) \chi_{m_z}(\phi) 
\end{equation}
are the real spherical harmonics, with $\Pjm$ being the associated Legendre functions, and 
\begin{equation}
\chi_{m_z} = \left\lbrace \begin{array}{lcc}
\dfrac{\sin(m_z\phi)}{\sqrt{\pi}} & \text{for} & m<0 \\
\dfrac{1}{\sqrt{2\pi}} & \text{for} & m=0 \\
\dfrac{\cos(m_z\phi)}{\sqrt{\pi}} & \text{for} & m>0
\end{array} \right.
\end{equation} 
being an azimuthal function; finally, the electric permittivity function is
\begin{equation}
\label{epsilon}
\epsilon(r)= \left\lbrace \begin{array}{lcc}
n_r^2 & \text{for} & r \leq R \\
1 & \text{for} & r > R \\
\end{array} \right.
\end{equation}
Notice that a constant magnetic permeability $\mu=1$ is assumed both for the embedding medium and the sphere. 

\subsection{Analysis of the effective volume}
Here, we show that the effective volumes of Mie resonators can be related to the translation-addition coefficients $\left\lbrace t_{j',m'_z}^{(e)},t_{j',m'_z}^{(m)}   \right\rbrace$ and the coefficients $\left\lbrace a_{j'},b_{j'} \right\rbrace$ (see Appendix). First of all, we re-write the Purcell factor formula given by eq.(\ref{Fe}) using the normal modes defined in eqs.(\ref{Ejme}-\ref{epsilon}) and the eigenfrequencies given by eqs.(\ref{oe}-\ref{om}):
\begin{equation}
F(\omega) = \dfrac{3\pi c^3}{\omega} \sum_{j,m_z,q} \left[   \mathrm{Im} \left\lbrace \dfrac{1}{\Ve \ome \left( \ome - \omega \right)} \right\rbrace + \mathrm{Im} \left\lbrace \dfrac{1}{\Vm \omm \left( \omm - \omega \right)} \right\rbrace \right]
\label{PFsph}
\end{equation}
where 
\begin{equation}
\Vx = \dfrac{1}{\left( \mathbf{u_p} \cdot \Ex ( \mathbf{r_p}, \omx   )\right)^2 }
\label{Veffgen}
\end{equation}
That is, ten parameters need to be specified to compute the effective volume for a dipolar emitter using eq.(\ref{Veffgen}): $\Vmu=V_{j,m_z,q}^{(x)}(\mathbf{r_p},\mathbf{u_p})$.

Now, it is enlightening to compare eq.(\ref{PFsph}) with the classical decay rates of the same system, which can be found in \cite{Chew1987}:
\begin{equation}
\dfrac{P}{P_0} = \dfrac{3c^6}{8\pi \vert p\vert^2 \omega^6}\sum_{j,m_z} \left[ \vert \ace + a_j \ace \vert^2 + \vert \acm + b_j \acm \vert^2 \right]
\end{equation}
The multipolar coefficients $\ace$ and $\acm$ defined in \cite{Chew1987} modulate the dipolar field exactly in the same manner as the translation-addition coefficients $\left\lbrace t_{j,m_z}^{(e)},t_{j,m_z}^{(m)}   \right\rbrace$ do in eq.(\ref{dip_multi}) \cite{Chew1987}. They are defined as:
\begin{eqnarray}
t_{j,m_z}^{(e)} = \ace \propto \mathbf{u_p} \cdot \Ae(\mathbf{r_p,\omega}) \\
t_{j,m_z}^{(m)} = \acm \propto \mathbf{u_p} \cdot \Am(\mathbf{r_p,\omega})
\end{eqnarray}
Hence, by comparison, we observe that the effective volumes can be analyzed thanks to the translation-addition coefficients:
\begin{equation}
\vert t_{j,m_z}^{(e)}(\omega) + a_j(\omega) t_{j,m_z}^{(e)}(\omega) \vert^2 \longleftrightarrow \sum_q \text{Im} \left\lbrace \dfrac{1}{\Ve\ome \left( \ome - \omega   \right)}    \right\rbrace
\label{compa}
\end{equation}
It is observed that the Purcell factor dependence on the azimuthal number $m_z$ is fully captured by the effective volume. For instance, if $t_{j,m'_z}^{(e)}(\omega)=0  \Longrightarrow  V_{j,m'_z,q}^{(e)} \rightarrow \infty$. But the dependence on $j$ is more complex, and so is the dependence on $\omega$, which is given by a summation of the kind $\sum_q 1/(A_q-B_q\omega)$. Thus, the effective volume quantifies the coupling between the electromagnetic field created by the emitter and the normal mode in consideration, as well as their spectral overlap.

\section{Purcell factor of silicon nano-cavities}
For the rest of the study, we consider a silicon-made Mie resonator with a refractive index $n_r=3.5$, magnetic permability $\mu=1$, of radius $R=0.2\mu m$, embedded in air. The normal modes of the system and their eigenfrequencies are computed using the formulas given by eqs.(\ref{oe}-\ref{epsilon}). It will be shown that even in the more symmetric situations, the Purcell factor has contributions from more than one normal mode. Also, we will show that, thanks to symmetry considerations, it is possible to predict the value of $\Vx$ for some cases. Finally, we will show that given a multipolar order $j'$, the average electric effective volume (it is defined next) is always smaller than the average magnetic effective volume. 

\subsection{Dipole in the center of the Mie resonator}
\begin{figure}[htbp]
\centering\includegraphics[width=13cm]{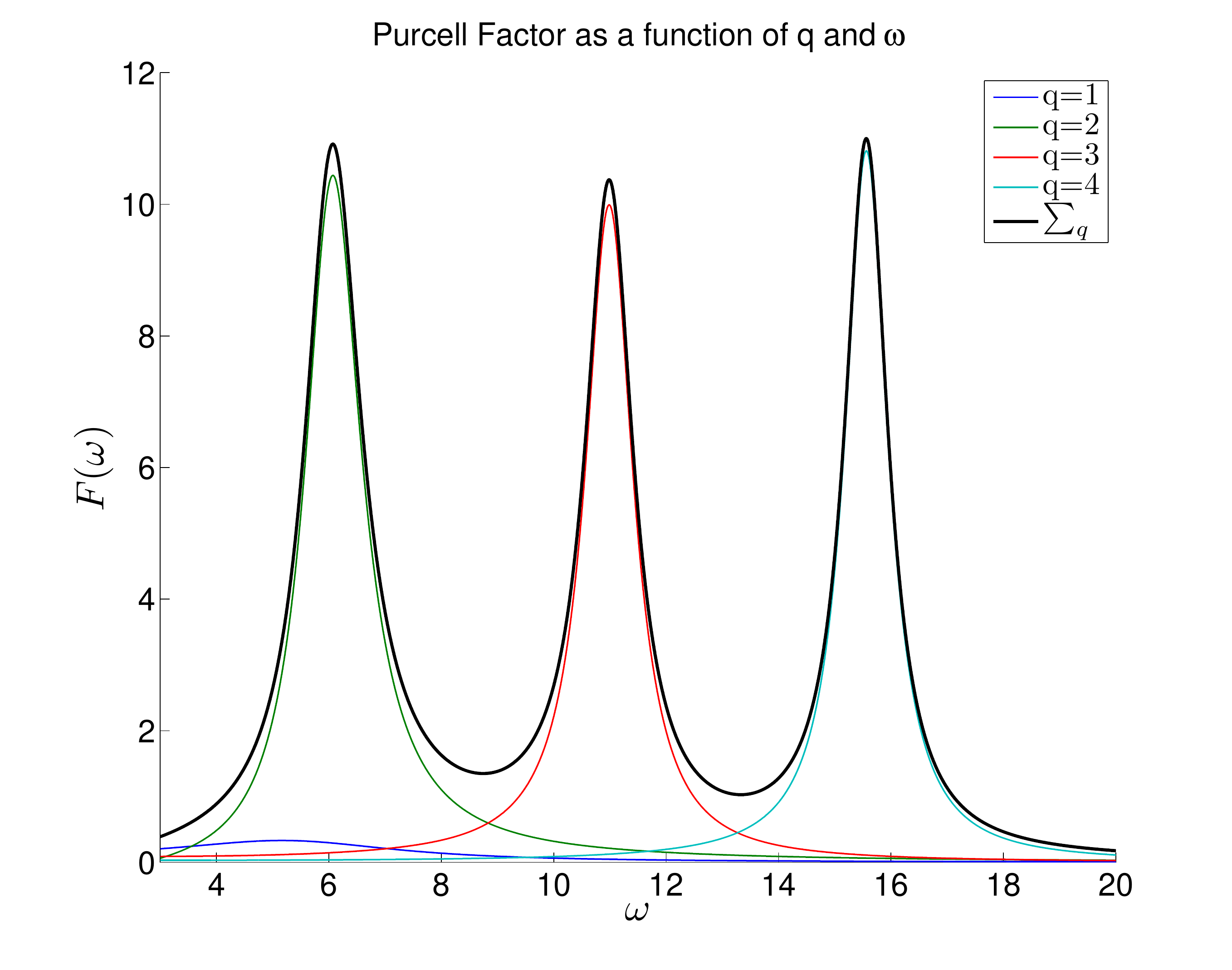}
\caption{Purcell factor ($F(\omega)$) of an electric dipole oriented along the $z$ axis located in the center of the resonator. Both the total Purcell factor and its partial $q$-contributions shown by eq.(\ref{Fq}) are depicted.}
\label{dipole_0}
\end{figure}

An electric dipole is first supposed to be placed at the origin of the resonator, oriented along the $z$ axis, \textit{i.e.} $\mathbf{p}=p\zhat$. Let us remind that the emitting frequency does not need to be given to compute the effective volumes. Thus, the electromagnetic field emitted by this dipole can be expressed simply as $\mathbf{E}^{\mathrm{d}}=\mathbf{A}_{1,0}^{(e)}$, \textit{i.e.} a single multipole (see appendix). That implies:

\begin{equation}
\begin{array}{ccccl}
t_{j,m_z}^{(m)} &=&0    \ , & \ & \forall \left( j,m_z \right)\\
t_{j,m_z}^{(e)}  &=&0   \ ,& \ &  \forall \left( j,m_z \right) \neq (1,0)
\end{array}
\end{equation}
Then, using eq.(\ref{compa}), it is observed that 
\begin{equation}
\begin{array}{ccccl}
\Vm & \rightarrow & \infty \ ,& \ & \forall(j,m_z,q) \\
\Ve & \rightarrow & \infty \ ,& \ & \forall(j,m_z,q) \neq (1,0,q)
\end{array}
\end{equation}
Thus, the Purcell factor is given by:
\begin{equation}
\label{Fq}
F(\omega)=\dfrac{3\pi c^3}{\omega}\sum_{q=1}^{\infty} \mathrm{Im} \left\lbrace \dfrac{1}{V_{1,0,q}^{(e)} \omega_{1,q}^{(e)} \left(\omega_{1,q}^{(e)} - \omega \right)} \right\rbrace
\end{equation}
In Table \ref{tab1}, we give the first four effective volumes and frequencies corresponding to $q=1,..,4$ units of $c=1$.
\begin{table}
\caption{\label{tab1}Effective volumes and eigenfrequencies associated to modes of the kind $\mathbf{E}_{1,0,q}^{(e)}$, for $q=1,..,4$. The effective volumes ($V_{1,0,q}^{(e)}$) are computed with respect to an electric dipole oriented along the $z$ axis and located in the center of the Mie resonator, using eq.(\ref{Veffgen}). The resonator is embedded in air, has an index of refraction $n_r=3.5$, and a radius $R=0.2 \mu m$. The eigenfrequencies ($\omega_{1,q}^{(e)}$) associated to $\mathbf{E}_{1,0,q}^{(e)}$ are computed with eq.(\ref{oe}).}
\begin{ruledtabular}
\begin{tabular}{ccccc}
 & $q=1$& $q=2$ & $q=3$ & $q=4$\\
\hline 
$V_{1,0,q}^{(e)}$ & $0.2514 + 0.2494i$ & $0.0383 + 0.0111i $ & $0.0151 + 0.0011i$ & $ 0.0077 + 0.0004i$ \\
\hline
$\omega_{1,q}^{(e)}$  &  $5.2940 - 2.4525i$ &   $6.0470 - 0.6070i$ & $10.9925 - 0.5150i$ & $15.5635 - 0.4665i$\\
\end{tabular}
\end{ruledtabular}
\end{table}
Using these values, the Purcell factor $F(\omega)$ as a function of the frequency of the emitter ($\omega$) is displayed in Fig.(\ref{dipole_0}). As it can be observed, $F(\omega)$ has contributions from different modes. When one of them is resonant, the behavior can almost be described with only one mode, but when they are not, this is no longer true. In addition, as it was shown in \cite{Koenderink2010}, when the Q-factor of a mode is not very high (as it is the case for $\mathbf{E}_{1,0,1}^{(e)} $), the Purcell factor can be underestimated if only one mode is considered.

\subsection{Dipole displaced along the z axis}\label{subseczaxis}
Here, we consider the case where an electric dipole with $\mathbf{p}=p\zhat$ is translated from the center of the resonator to its surface along the $z$ axis. As discussed before, the electric field emitted by the dipole from a frame of reference centered on itself is given by $\mathbf{E}^{\mathrm{d}}=\mathbf{A}_{1,0}^{(e)}$. However, the description of the normal modes of the cavity with respect to a frame of reference centered at the displaced dipole becomes cumbersome. It is easier to describe $\mathbf{E}^{\mathrm{d}}$ from the frame of reference centered at the center of the cavity. Therefore, the translation-addition coefficients $t_{j,m_z}^{(x)}$ are used to express $\mathbf{E}^{\mathrm{d}}$ with respect to a different point $z_0$ on the $z$ axis. As we showed in eq.(\ref{dip_multi}), the expression is a general superposition of multipolar fields. Now, we demonstrate that a translation along the $z$ axis restricts this general superposition to a case where $t_{j,m_z}^{(e)}=f(j)\delta_{0,m_z}$ and therefore:
\begin{equation}
\Ed=\mathbf{A}_{1,0}^{(e)}(\mathbf{r}-z_0 \zhat)=\sum_{j'} t_{j',0}^{(e)} \mathbf{A}_{j',0}^{(e)}
\end{equation}
On one hand, the multipolar fields are eigenstates of the $z$ component of the angular momentum, $J_z$, \textit{i.e.} $J_z \mathbf{A}_{j,m_z}^{(x)} = m_z\mathbf{A}_{j,m_z}^{(x)} $ \cite{Zambrana2013JQSRT,ZambranaThesis,Tung1985}. On the other hand, a translation along the $z$ axis can be expressed as a function of the $z$ component of the linear momentum, $P_z$: $T_z(\Delta z)=\exp\left( -i P_z \Delta z  \right)$. Then, because $J_z$ and $P_z$ commute, $\left[ J_z, P_z   \right]=0$, a translation along the $z$ axis must maintain the $z$ component of the angular momentum \cite{Sakurai1995,Tung1985}. A justification why $t_{j,m_z}^{(m)}=0$ can be found in \cite{Tung1985}. With the relations brought forward by eq.(\ref{compa}), we obtain that:
\begin{equation}
\begin{array}{ccccl}
\Vm & \rightarrow & \infty  \ ,& \ & \forall(j,m_z,q) \\
\Ve & \rightarrow & \infty  \ ,& \ & \forall(j,m_z,q) \neq (j,0,q)
\end{array}
\end{equation}
Hence, the Purcell factor can be analytically written in this occasion as:
\begin{equation}
F(\omega)=\dfrac{3\pi c^3}{\omega}\sum_{j,q} \mathrm{Im} \left\lbrace \dfrac{1}{V_{j,0,q}^{(e)} \omega_{j,q}^{(e)} \left(\omega_{j,q}^{(e)} - \omega \right)} \right\rbrace
\end{equation}
Notice that even though the formula is still very simplified (with respect to eq.(\ref{PFsph})), the number of mode volumes and eigenfrequencies that need to be computed in order to calculate $F(\omega)$ increases a lot. In Table \ref{tab2}, we give the eigenfrequencies of all the normal modes whose frequencies lay in the interval given by the frequencies given in Table \ref{tab1}, \textit{i.e.} $\mathrm{Re} \left\lbrace \omega_{j,q}^{(e)} \right\rbrace \in [4, 16]$.
\begin{table}
\caption{\label{tab2}Eigenfrequencies of a Mie resonator embedded in air, with $n_r=3.5$, and $R=0.2 \mu m$. Only the frequencies in interval $\mathrm{Re} \left\lbrace \omega_{j,q}^{(e)} \right\rbrace \in [4, 16]$ are listed.}
\begin{ruledtabular}
\begin{tabular}{ccccc}
 & $q=1$& $q=2$ & $q=3$ & $q=4$\\
\hline 
$\omega_{1,q}^{(e)}$  &  $5.2940 - 2.4525i$ &   $6.0470 - 0.6070i$ & $10.9925 - 0.5150i$ & $15.5635 - 0.4665i$\\
\hline
$\omega_{2,q}^{(e)}$  &  $7.7377 - 0.1466i$ &   $10.5360 - 3.3610i$ & $12.7485 - 0.6925i$ & \\
\hline
$\omega_{3,q}^{(e)}$  &  $9.6137 - 0.02975i$ &   $14.2235 - 0.4400i$ & $15.8325 - 3.8870i$ & \\
\hline
$\omega_{4,q}^{(e)}$  &  $11.392145 - 0.0059795i$ &   $16.1388499 - 0.138451i$ &  & \\
\hline
$\omega_{5,q}^{(e)}$  &  $13.1071675 - 0.0011945i$ &   &  & \\
\hline
$\omega_{6,q}^{(e)}$  &  $14.78344639 - 0.0002351i$ &    &  & 
\end{tabular}
\end{ruledtabular}
\end{table}
\begin{figure}[htbp]
\centering\includegraphics[width=13cm]{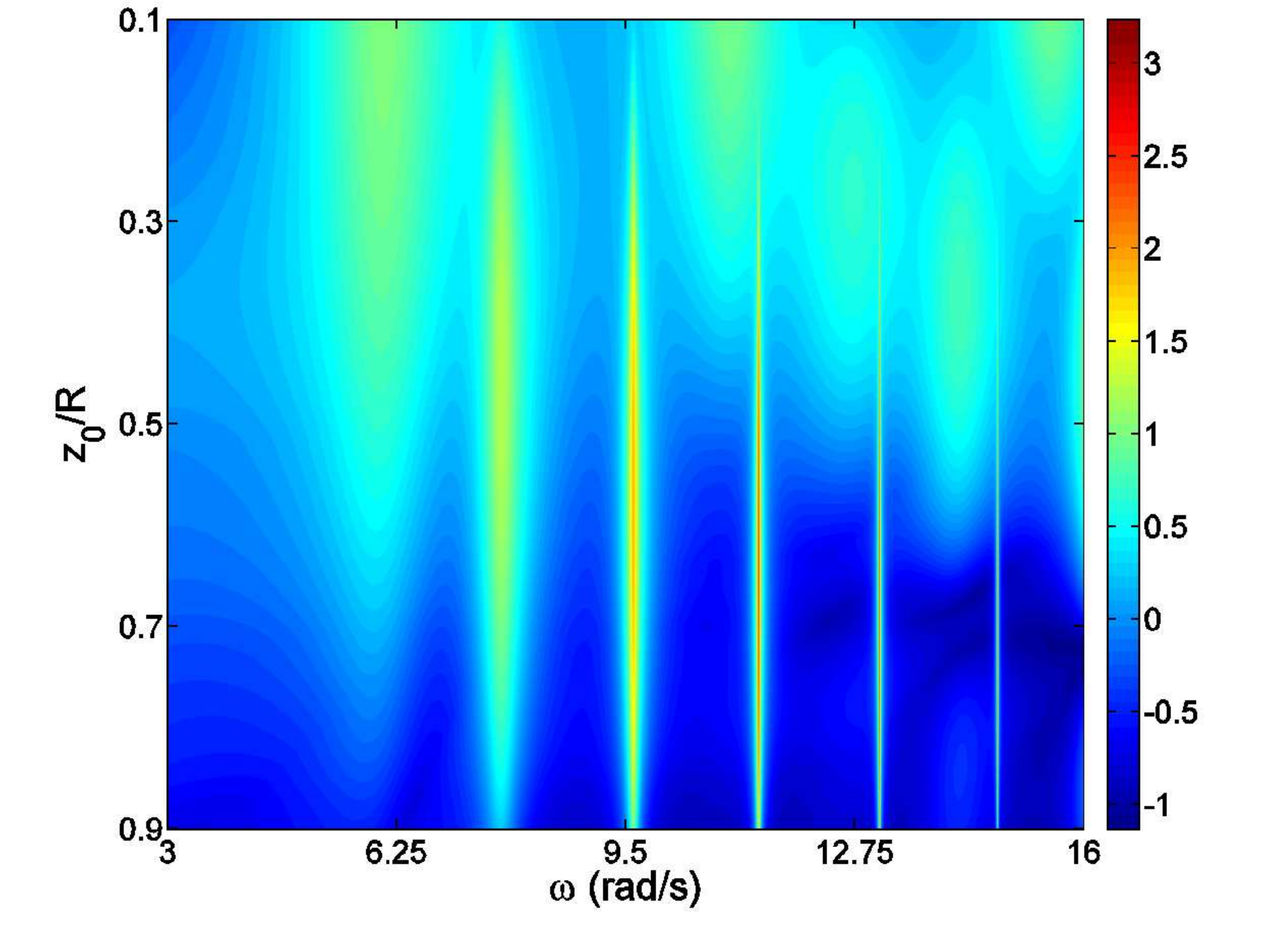}
\caption{Logarithm of the Purcell factor ($\log \left[ F(\omega,z_0) \right]$) as a function of the frequency $\omega$ of the emitter and its position $z_0$ in the $z$ axis. The orientation of the dipole is along the $z$ axis.  The Mie resonator is embedded in air and is defined by $n_r=3.5$, and $R=0.2 \mu m$.}
\label{PFz0}
\end{figure}
In Fig.\ref{PFz0}, we plot $\log\left( F(\omega,z_0)  \right) $, \textit{i.e.} the Purcell factor as a function of the emitting frequency of the dipole ($\omega$) and its position in the $z$ axis ($z_0$). 
\begin{figure}[htbp]
\centering\includegraphics[width=13cm]{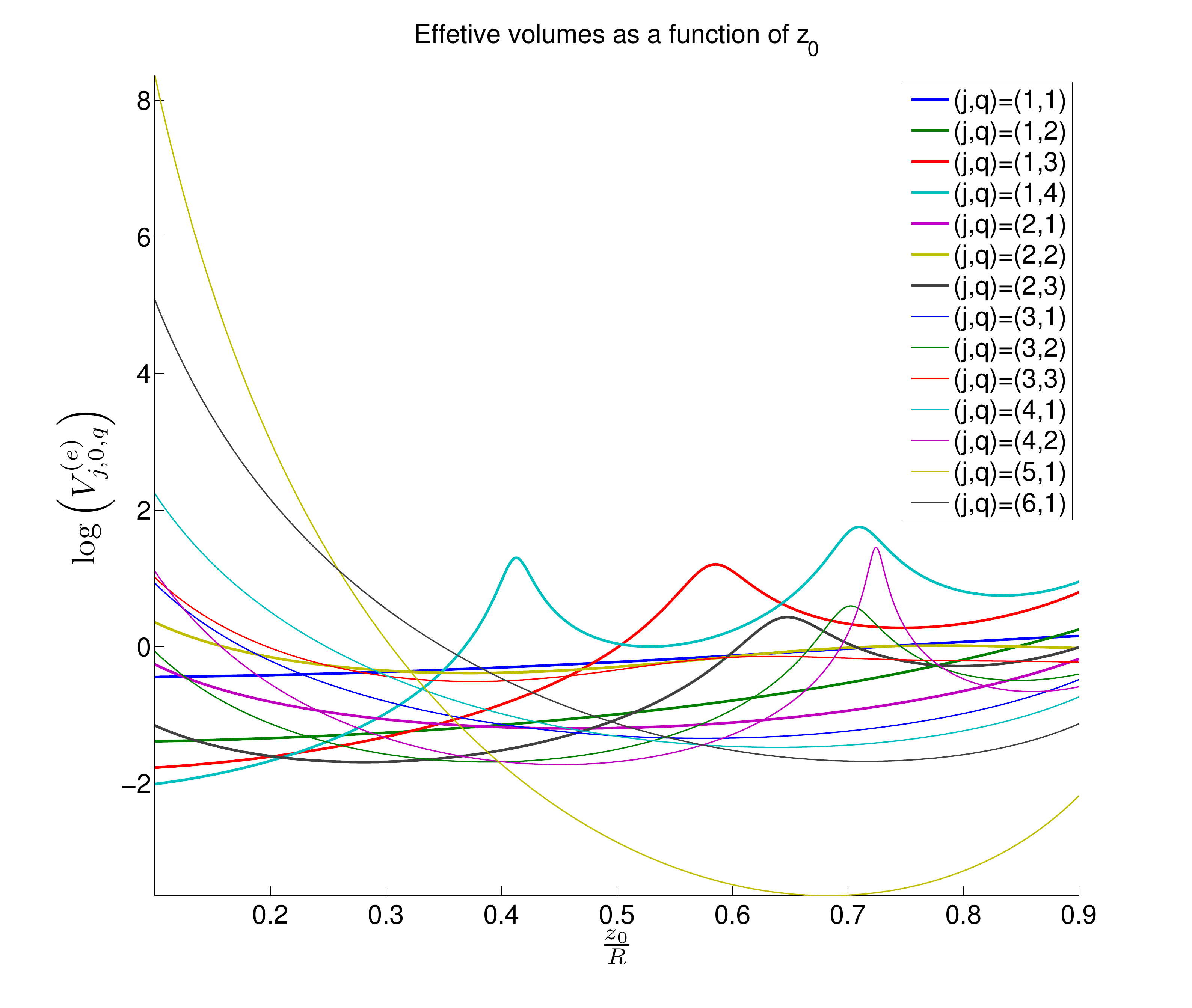}
\caption{Logarithm of the effective volumes ($\log\left( V_{j,0,q}^{(e)} \right)$) of normal modes of the kind $\mathbf{E}_{j,0,q}^{(e)}$ as a function of the position of the emitter $z_0$. Only the normal modes associated to the eigenfrequencies given by Table \ref{tab2} are considered. The emitter is an electric dipole oriented along the $z$ axis, and its position is moved along the $z$ axis. The Mie resonator is embedded in air and is defined by $n_r=3.5$, and $R=0.2 \mu m$.  }
\label{EffVolz0}
\end{figure}
The logarithmic function is used to visualize the behavior of $\log\left( F(\omega,z_0)  \right) $.

\begin{figure}[htbp]
\centering\includegraphics[width=16cm]{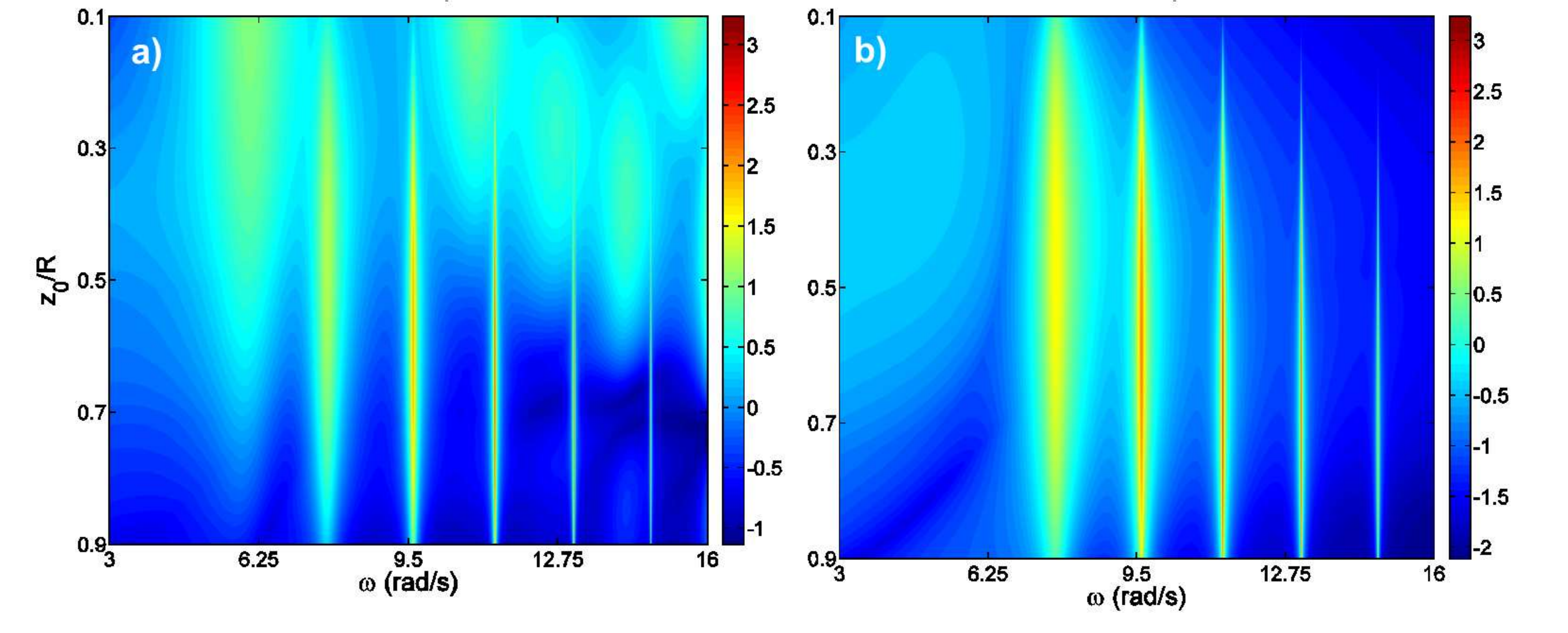}
\caption{Comparison between a) the Purcell factor computed using all the normal $q$-modes of the resonator and b) the Purcell factor computed only considering modes with $q=1$. In both cases, the Purcell factor is computed as a function of the frequency $\omega$ of the emitter and its position $z_0$ in the $z$ axis. The emitter is oriented along the $z$ axis in both cases. The Mie resonator is consideration is embedded in air and is defined by $n_r=3.5$, and $R=0.2 \mu m$.  }
\label{fig5}
\end{figure}

It is observed that when the dipole is very close to the origin, $\log\left( F(\omega,z_0)  \right) $ has broad resonances, as obtained in Fig.\ref{dipole_0}. However, when the dipole is displaced from the origin, its coupling to the resonator gets stronger as well as more resonant. Note that even though some of the resonances have an extremely narrow linewidth, their contribution to the Purcell factor is not present for all the positions of the dipole. For example, $\omega_{5,1}^{(e)}$ almost does not contribute to the Purcell factor for $\frac{z_0}{R} < 0.3 $. The reason why this happens is that the effective volume $V_{5,0,1}^{(e)}$ is very large for $\frac{z_0}{R} > 0.3 $. This can be observed in Fig.\ref{EffVolz0}, where the effective volumes of all the modes used to compute $F(\omega,z_0)$ in Fig.\ref{PFz0} are given. Finally, note that even though the most resonant behavior is given by the modes with $q=1$, the whole Purcell factor picture would be different if only modes with $q=1$ were included. A comparison between the two has been depicted in Fig.\ref{fig5}.

\subsection{Effective volumes. Average values}
In this section, we describe the computation of average effective volumes and show that given a multipolar order $j$, the average effective volumes are always smaller for electric modes. Suppose that a Mie resonator is doped with a large number of dipolar emitters. In general, both their locations and orientations in the resonator are unknown. Thus, we can define the average effective volume as:
\begin{equation}
\left\langle \Vx \right\rangle = \int_0^R \int_{\Omega} \dfrac{3r^2 \dd r \dd \Omega}{[\rhat \cdot \Ex]^2+[\that \cdot \Ex]^2+[\phat \cdot \Ex]^2}
\label{Aveffvol}
\end{equation}
Because the order of the integrals can be interchanged, we can also define an average angular effective volume:
\begin{equation}
 \left\langle \Vx (r) \right\rangle = \int_{\Omega} \dfrac{3 \dd \Omega}{[\rhat \cdot \Ex(r)]^2+[\that \cdot \Ex(r)]^2+[\phat \cdot \Ex(r)]^2}
\end{equation}
so that the average effective volume is computed as:
\begin{equation}
\left\langle \Vx \right\rangle = \int_0^R r^2 \left\langle \Vx (r) \right\rangle \dd r 
\end{equation}
Note that in this case, there is only one symmetry argument, which is that $V_{j,m_z,q}^{(x)}=V_{j,-m_z,q}^{(x)}$. The reason stems from the definition of the normal modes of the systems given by eqs.(\ref{Ejme}-\ref{epsilon}). In particular, the only dependence of $\Ex$ in $m_z$ comes via $Y_{jm_z}$ and $\chi_{m_z}$, and it can be seen that an integral over $2\pi$ equalizes the behavior of the modes with $m_z$ and $-m_z$. Thus, in general, the average effective volume $\aVx$ does not tend to infinity for all $j,m_z,q,(x)$. In Table \ref{AvVol} we present the values of the inverse of $\aVx$ for different electric and magnetic modes with $q=1$. If $1/\aVx$ is large, the mode in consideration contributes significantly to the Purcell factor at $\omega = \mathrm{Re} \left\lbrace \omx \right\rbrace$. That is, a large value of $1/\aVx$ means good coupling between that mode and the resonator. 
\begin{table}
\caption{\label{AvVol}Computation of $1/\aVx$ for different normal modes of the resonator. All the volumes have been computed with $q=1$. The variation in $\left\lbrace j,m_z,(x) \right\rbrace$ is made explicit in the table. Large values of $ \vert 1/\aVx \vert$ indicate good coupling between the uniform distribution of dipoles in the resonator and the mode in consideration. The resonator is embedded in air and is defined by $n_r=3.5$, and $R=0.2 \mu m$. }
\begin{center}
\begin{tabular}{|c|c|c|c|}
\hline $j$& $m_z$ & $(e)$ & $(m)$ \\
\hline \multirow{2}{*}{$j=1$} & $m_z=0$ & $22.6754 -12.5357i $ & $0.778488  -0.0930585i$ \\
&  $m_z=1$ & $22.6754  -12.5359i$ & $4.69505 + -0.561234i$  \\
\hline \multirow{3}{*}{$j=2$} & $m_z=0$ & $8.86982  -0.944547i$ & $0.0222563 + -0.0020265i$ \\
&  $m_z=1$ & $2.89243  -0.745228i $ & $0.310968  -0.0283146i$  \\
&  $m_z=2$ & $0.674872 + -1.2691i $ & $0.0814357  -0.00741497i$  \\
\hline \multirow{4}{*}{$j=3$} & $m_z=0$ & $4.39869  -0.0542868i $ & $2.14545\e{-5}  -6.63047\e{-7}i$ \\
&  $m_z=1$ & $0.0743161 + 0.120752i $ & $0.000216611  -6.69432\e{-6}i$  \\
&  $m_z=2$ & $-0.157549 + 0.0998519i$ & $0.000122304  -3.7798\e{-6}i$  \\
&  $m_z=3$ & $-8.03716\e{-10} + 2.58389\e{-10}i $ & $1.60496\e{-12}  -4.96009\e{-14}i$  \\
\hline
\end{tabular}
\end{center}
\end{table}
Two observations can be made. Firstly, it is seen that the coupling $1/\aVx$ decreases with the multipolar order $j$. This result can be interpreted using the radial dependence of multipolar fields. Indeed, the intensity of a multipolar field has a radial dependence that is given by a spherical Bessel function $j_j(kr)$. Now, when given a fixed $kr^*$ position close to the origin, $j_{j-1}(kr^*) >j_{j}(kr^*)$ \citep{Abramowitz1970}. That is, lower multipolar orders have a non-null intensity in the center of the sphere, whereas the intensity of higher orders is pushed to the surface. Consequently, smaller values of $\vert \Ex \vert$ contribute to the integral given by eq.(\ref{Aveffvol}) for most of the points in the domain. 

Secondly, given a multipolar order $j$, the coupling of the magnetic modes is always smaller than the coupling to the electric ones. This fact can be understood using eq.(\ref{compa}). As we saw for the case of a displaced dipole along the $z$ axis, the coupling of a dipolar emitter to magnetic modes is completely suppressed. This fact changes when the dipoles are not located or oriented along the $z$ axis, but the coupling with electric modes (given by the translation-addition coefficients $t_{j,m_z}^{(e)}$) still remains stronger. 

\begin{figure}[htbp]
\centering\includegraphics[width=13cm]{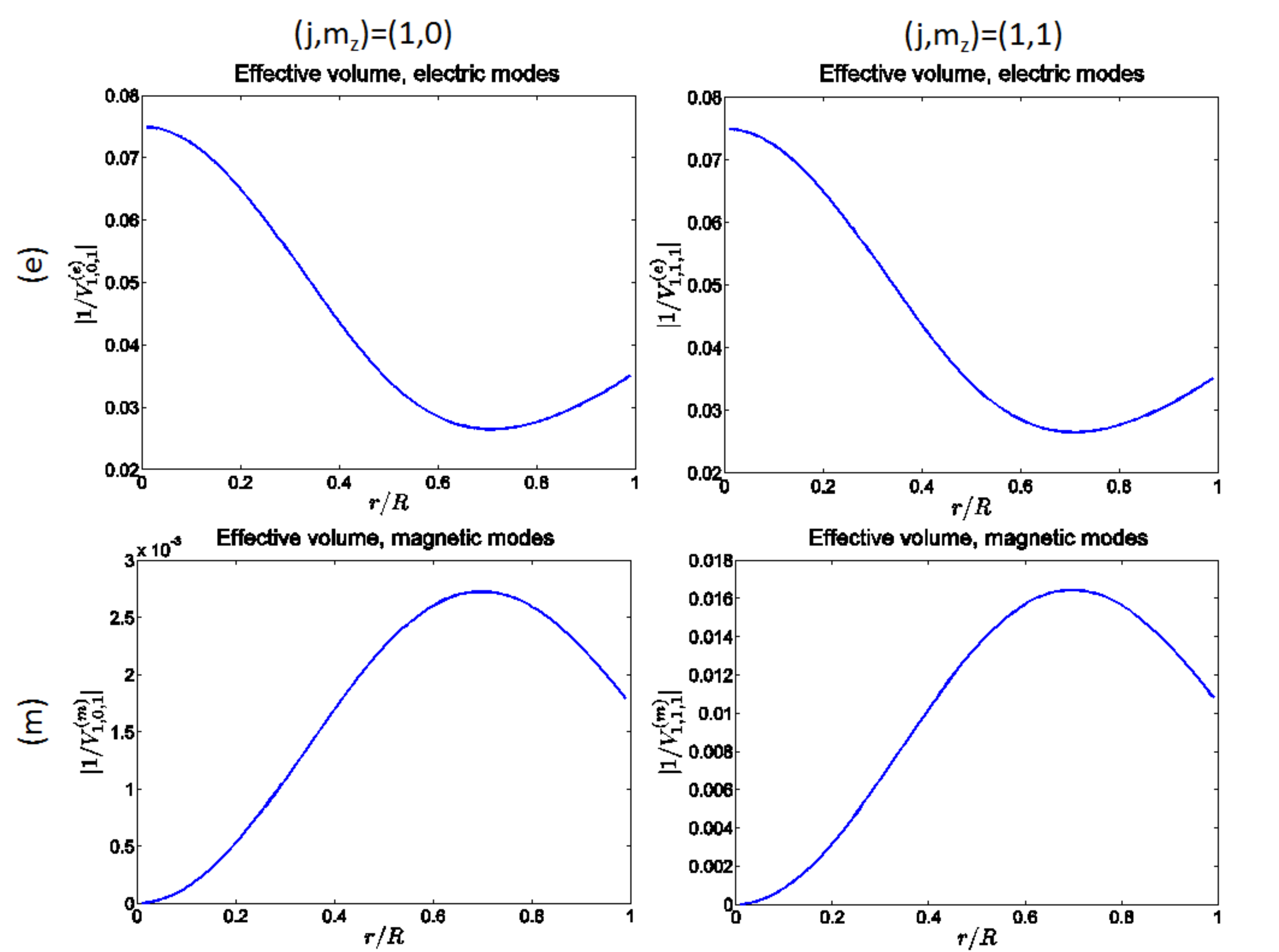}
\caption{Absolute values of the inverse of the angular average effective volume ($ \vert 1/ \left\langle \Vx (r) \right\rangle \vert $ ) for $j=1$ and $q=1$. Large values of $\vert 1/ \left\langle \Vx (r) \right\rangle \vert $ indicate good coupling between the uniform distribution of dipoles in the resonator and the mode in consideration. The resonator is embedded in air and is defined by $n_r=3.5$, and $R=0.2 \mu m$. }
\label{j1}
\end{figure}

\begin{figure}[htbp]
\centering\includegraphics[width=16cm]{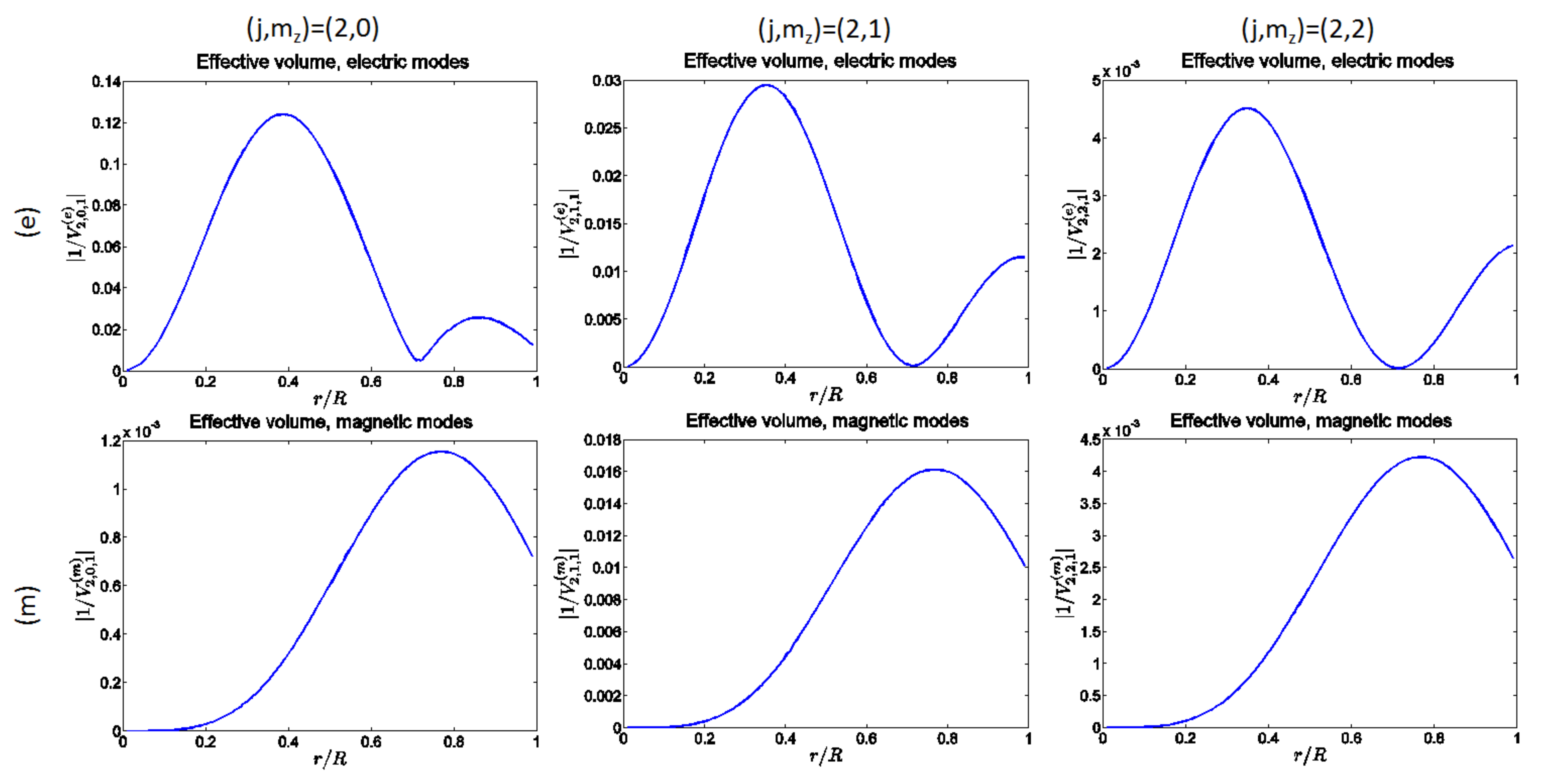}
\caption{Absolute value of the inverse of the angular average effective volume ($ \vert 1/ \left\langle \Vx (r) \right\rangle \vert $ ) for $j=2$ and $q=1$. Large values of $\vert 1/ \left\langle \Vx (r) \right\rangle \vert $ indicate good coupling between the uniform distribution of dipoles in the resonator and the mode in consideration. The resonator is embedded in air and is defined by $n_r=3.5$, and $R=0.2 \mu m$.}
\label{j2}
\end{figure}

Finally, we plot  $1/ \left\langle \Vx (r) \right\rangle $ for a few normal modes. This is depicted in Fig.\ref{j1} and Fig.\ref{j2}. Note that the results of Table \ref{AvVol} are computed as radial integrals of these plots.  
\section{Conclusions}
In conclusion, we derive the Purcell factor for magnetic and electric dipolar emitters. We particularize it for the case of spherical Mie resonators and emitting electric dipoles. Using the eigenfrequencies and the analytic expressions of the normal modes of the system, we analyze the meaning of the effective volume for spherical cavities. It is seen that its value is related to the multipolar coefficients in the translation-addition formulas. That is, the effective volume of a normal mode locally quantifies the strength of the coupling of a dipolar emitter to that normal mode. Finally, the Purcell factor is computed for different symmetric scenarios. We observe that the magnetic effective volumes are $\Vm \rightarrow \infty$ for some symmetric configurations, showing no coupling between the electric emitter and the magnetic modes of the structure. However, in less symmetric configurations, this is not necessarily the case. Finally, an average effective volume is computed both for electric and magnetic modes, revealing that $\aVe < \aVm$, \textit{i.e.} the coupling to electric modes is generally better than the coupling to magnetic modes. The situation could be reverted if the effective volumes were computed with dipolar magnetic emitters.

\section{acknowledgments}
The authors want to thank Alexis Devilez, Brian Stout, Jean-Paul Hugonin, Christophe Sauvan, Philippe Lalanne and G\'{e}rard Colas des Francs for fruitful discussions. This work has been carried out thanks to the support of the A*MIDEX project (n$^{\circ}$ ANR-11-IDEX-0001-02) funded by the Investissements d'Avenir French Government program managed by the French National Research Agency (ANR).

  \renewcommand{\theequation}{A.\arabic{equation}}
  \setcounter{equation}{0}  
  \section*{Appendix}  
Here, we give describe the interaction of a dipolar emitter and a Mie resonator.
As it is described in \cite{Wittmann1988}, an emitting dipole is a multipolar field with $j=1$. If the dipole is electric, then its associated electric field can be described as $\mathbf{A}_{1,m_z}^{(e)}$ with respect to a frame of reference centred on the dipole itself, where $m_z$ depends on the orientation of the dipole. In fact, $m_z=\pm 1$ if the electric dipole is given by $\mathbf{p}=p \frac{\xhat \mp i \yhat}{\sqrt{2}}$, or $m_z=0$ if $\mathbf{p}=p \zhat$. Then, when $\mathbf{A}_{1,m_z}^{(e)}$ is expressed with respect to a reference of frame which at a distance $\mathbf{r}_0$ of it, the dipolar field is re-expressed as
\begin{equation}
\label{dip_multi}
\Ed=\mathbf{A}_{1,m_z}^{(e)}(\mathbf{r}-\mathbf{r}_0)=\sum_{j',m'_z} t_{j',m'_z}^{(e)} \mathbf{A}_{j',m'_z}^{(e)} + t_{j',m'_z}^{(m)} \mathbf{A}_{j',m'_z}^{(m)} 
\end{equation} 
with $\left\lbrace t_{j',m'_z}^{(e)},t_{j',m'_z}^{(m)}   \right\rbrace$ being the translation-addition coefficients, that can be found in \cite{Wittmann1988,Mishchenko2002}. The magnetic field can be computed using the relation \cite{Bohren1983}:
\begin{equation}
\label{H}
\mathbf{H}= i\dfrac{\nabla \times \mathbf{E}}{\omega \mu k}
\end{equation}
Due to the symmetries of the problem, the interior and scattered field are given by \cite{ZambranaThesis}:
\begin{eqnarray}
\label{Eint}
\Eint = \sum_{j',m'_z} d_{j'} t_{j',m'_z}^{(e)} \mathbf{A}_{j',m'_z}^{(e)} + c_{j'} t_{j',m'_z}^{(m)} \mathbf{A}_{j',m'_z}^{(m)} \\
\Es = \sum_{j',m'_z} a_{j'} t_{j',m'_z}^{(e)} \mathbf{A}_{j',m'_z}^{(e)} + b_{j'} t_{j',m'_z}^{(m)} \mathbf{A}_{j',m'_z}^{(m)}
\label{Es}
\end{eqnarray}
where $\{ a_{j'},b_{j'},c_{j'},d_{j'} \}$ are functions of the multipolar order $j$; the ratio between the index of refraction of the resonator and the embedding medium $n_r$; and the size parameter of the problem, given by $x= \frac{2\pi R}{\lambda}$, with $R$ the radius of the spherical resonator and $\lambda$ the wavelength of the emission of the radiating dipole. The coefficients ${a_j,b_j,c_j,d_j}$ are known as Mie coefficients when the dipole is outside of the resonator, but take a different form when the dipole is inside of it \cite{Chew1976,Chew1987}. The magnetic fields $\Hint$ and $\Hs$ are computed with eq.(\ref{H}).

\end{document}